\documentclass{aa}

\usepackage{graphicx}
\usepackage{times}
\usepackage{dcolumn}

\newcommand{\he}{\mbox{HE~0435$-$1223}}
\newcolumntype{.}{D{.}{.}{-1}}

\begin{document}


\title{HE~0435$-$1223: a wide separation quadruple QSO and gravitational lens
          \thanks{Based in part on observations obtained with the Baade 6.5-m
                  telescope of the Magellan Consortium.
                  Also based in part on observations collected at the European
                  Southern Observatory, La Silla, Chile.}}

\author{Lutz Wisotzki\inst{1} 
        \and 
        Paul L. Schechter\inst{2,3}
        \and
        Hale V. Bradt\inst{2}
        \and
        Janine Heinm\"uller\inst{1}
        \and
        Dieter Reimers\inst{4}
        }

\authorrunning{L. Wisotzki et al.}
\titlerunning{HE~0435$-$1223: a wide separation quadruple QSO}

\institute{%
           Universit\"at Potsdam, Am Neuen Palais 10, 14469 Potsdam, Germany, 
           email: lutz@astro.physik.uni-potsdam.de
           \and
           Dept.\ of Physics, Massachusetts Institute of
           Technology, Cambridge, MA 02139, USA
           \and
           Institute for Advanced Study, Einstein Drive, Princeton,
           NJ 08540, USA
           \and
           Hamburger Sternwarte, Universit\"at Hamburg, Gojenbergsweg 112, 
           21029 Hamburg, Germany
          }

\date{Draft \today}

\abstract{We report the discovery of a new gravitationally lensed QSO, 
at a redshift $z = 1.689$,
with four QSO components in a cross-shaped arrangement around a bright galaxy.
The maximum separation between images is $2\farcs 6$, enabling a 
reliable decomposition of the system. Three of the QSO components
have $g\simeq 19.6$, while component A is about 0.6~mag brighter. 
The four components have nearly identical colours, suggesting little
if any dust extinction in the foreground galaxy. The lensing galaxy
is prominent in the $i$ band, weaker in $r$ and not detected in $g$.
Its spatial profile is that of an elliptical galaxy with a scale length
of $\sim 12$\,kpc. Combining the measured colours and a mass model
for the lens, we estimate a most likely redshift range of $0.3 < z < 0.4$.
Predicted time delays between the components are $\la$ 10 days.
The QSO shows evidence for variability, with total $g$ band
magnitudes of 17.89 and 17.71 for two epochs separated by $\sim 2$ months.
However, the relative fluxes of the components did not change, 
indicating that the variations are intrinsic to the QSO rather than 
induced by microlensing. 
}

\maketitle

\section{Introduction}

The majority of known gravitationally lensed QSOs displays
image splitting into two components. Such systems offer relatively
few constraints for their mass distributions -- usually just
the two positions since the flux ratios might be 
changed by microlensing.
Both for the purpose of studying lensing potentials 
(Keeton, Kochanek \& Falco \cite{keeton*:98:OPGL})
and for the purpose of measuring cosmological parameters 
(Schechter \cite{schechter:01:GL}), 
quadruply imaged quasars are considerably more useful than 
their doubly imaged counterparts.
The discovery of a new quadruply imaged is therefore welcomed 
by everyone except those who struggle to explain why their 
relative numbers are so high in radio lensing surveys
(Rusin and Tegmark \cite{rusi+tegm:01:FFI}).

In this paper we report the discovery of a new gravitationally
lensed QSO with quadruple image splitting. The object was found
to be multiple as part of a new, ongoing imaging survey for lensed quasars
using the Magellan Consortium's 6.5~m Baade telescope on Cerro Las
Campanas. We present the first spectrum of this QSO and analyse
\emph{gri} imaging data to establish a first suite of astrometric
and photometric properties of the QSO components and the lensing
galaxy. We then discuss the available constraints on the surface
mass distribution in the lens and conclude with some prospects
for future observations.

\section{Observations}    \label{sec:obs}

\subsection{Identification and spectroscopy}    \label{sec:spec}

\begin{figure}[tb]
\includegraphics[width=8.8cm,bb=80 86 365 286,clip]{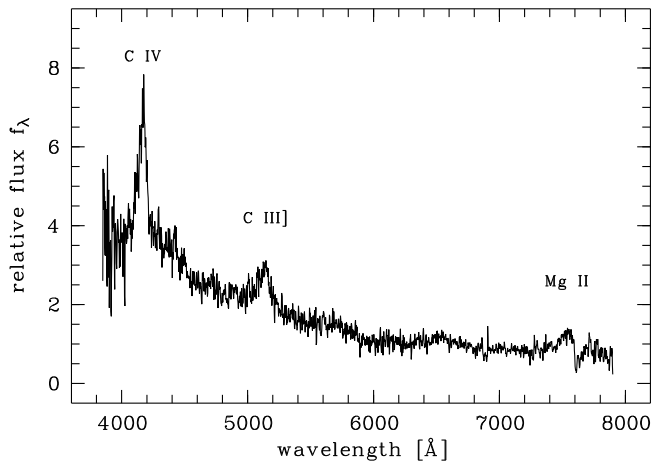}
\caption[]{Spatially unresolved slit spectrum of \he{},
   taken prior to the Magellan images in order
   to confirm the QSO nature of the source. }
   \label{fig:spec}
\end{figure}

\he{} was originally found as a high-probability QSO candidate,
selected in the course of the Hamburg/ESO digital objective prism survey 
(Wisotzki et al.\ \cite{wisotzki*:00:HES3}).
Subsequent low-resolution spectroscopy was performed in March 2000 
using the ESO 1.52~m telescope on La Silla with its Boller~\&~Chivens 
spectrograph. A 5~min exposure was enough to confirm the nature
of this object as a bona~fide QSO, and to determine its redshift 
$z=1.689$ (measured from the peak of the C\,{\sc iv} emission line).
The spectrum is displayed in Fig.~\ref{fig:spec} and reveals a fairly
average QSO without obvious peculiarities. At that time,
only a photographic sky survey image existed, showing the QSO as
an unresolved single source. It is now clear that only a fraction 
of the quadruple image ensemble was located within the $2''$ spectrograph
slit, and the spectrum therefore constitutes an ill-defined blend
of all four components. However, the very similar optical colours 
(see below) make it likely that Fig.\ \ref{fig:spec} is representative
for each of the QSO components.

\begin{figure*}[tb]
\setlength{\unitlength}{1mm}
\begin{picture}(180,124)
\put(0,84){\includegraphics[width=4cm,bb=98 98 302 302]{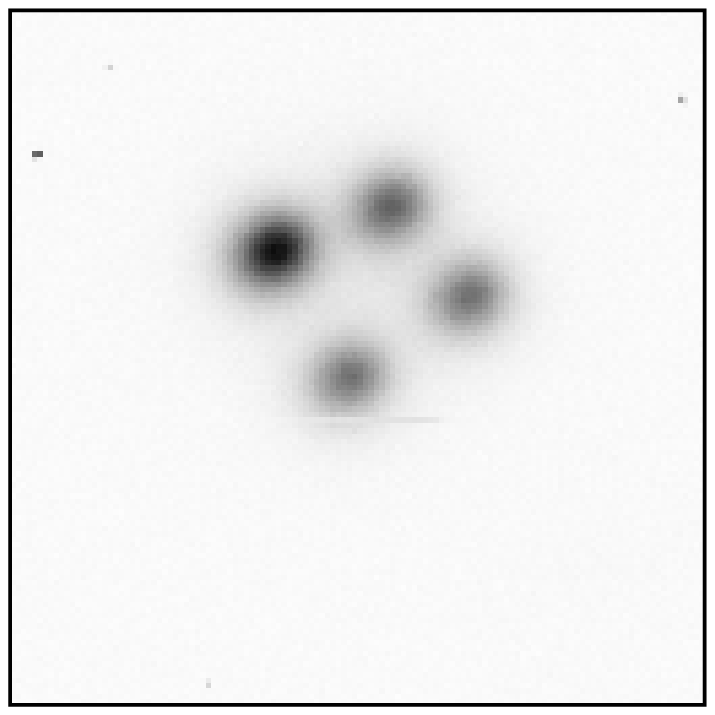}}
\put(41,84){\includegraphics[width=4cm,bb=98 98 302 302]{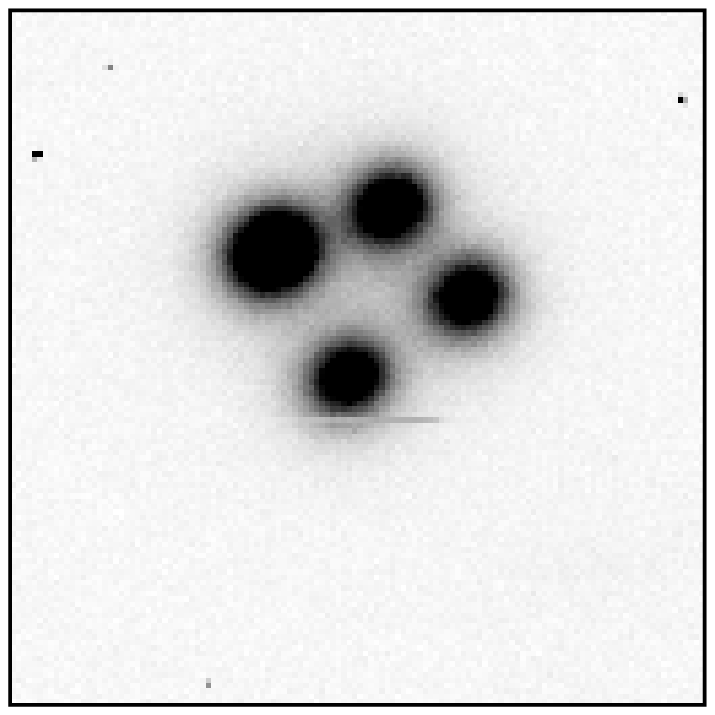}}
\put(82,84){\includegraphics[width=4cm,bb=98 98 302 302]{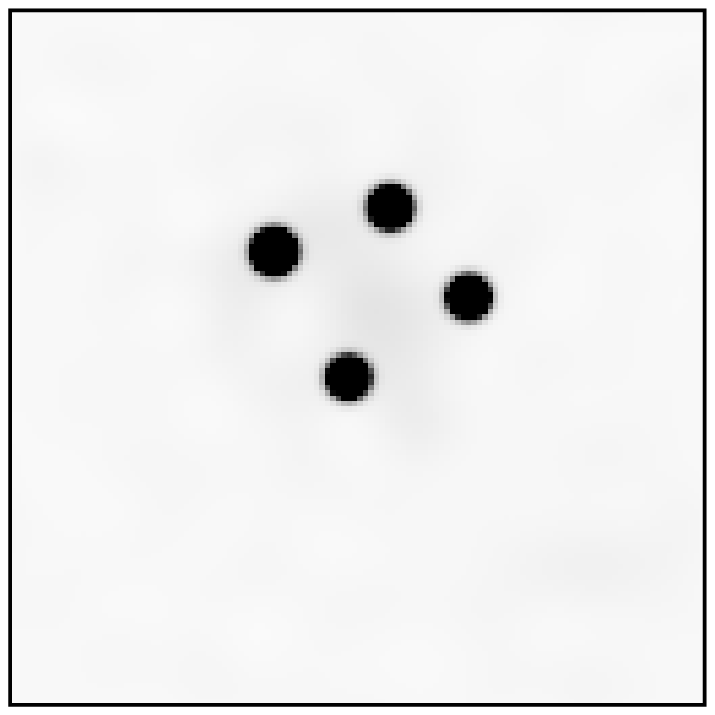}}
\put(123,84){\includegraphics[width=4cm,bb=98 98 302 302]{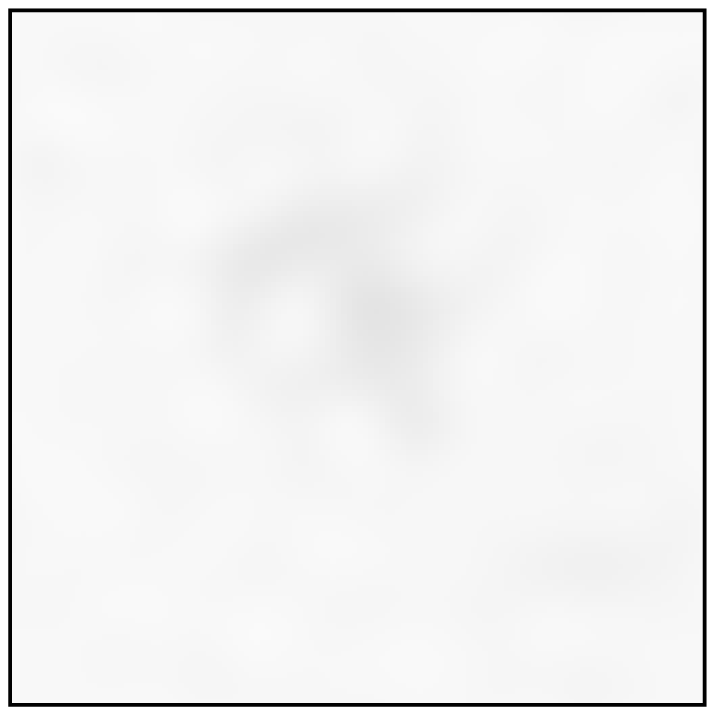}}
\put(5,90){\LARGE$g$}
\put(0,42){\includegraphics[width=4cm,bb=98 98 302 302]{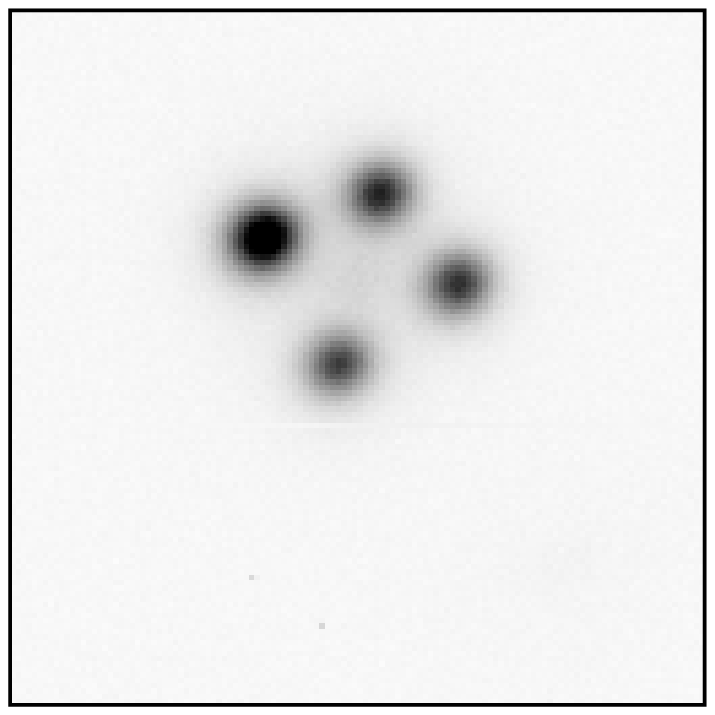}}
\put(41,42){\includegraphics[width=4cm,bb=98 98 302 302]{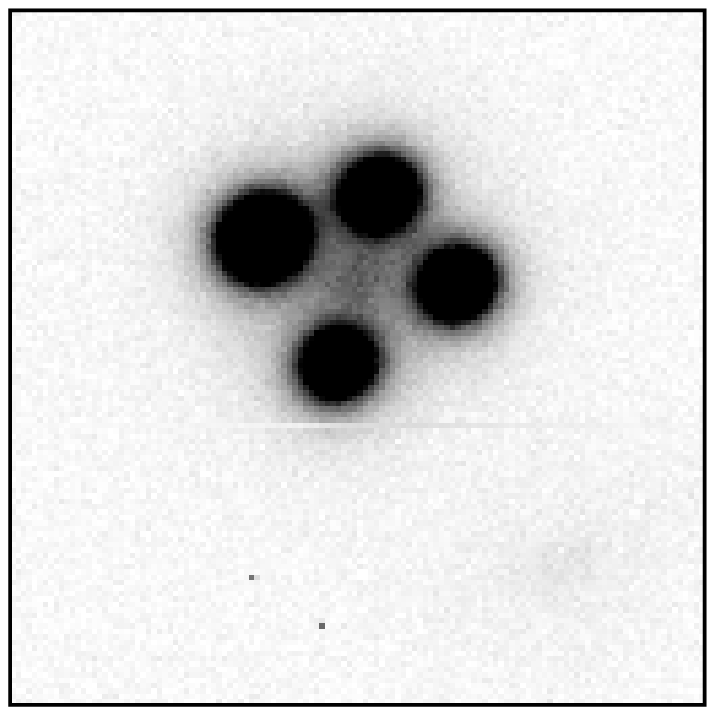}}
\put(82,42){\includegraphics[width=4cm,bb=98 98 302 302]{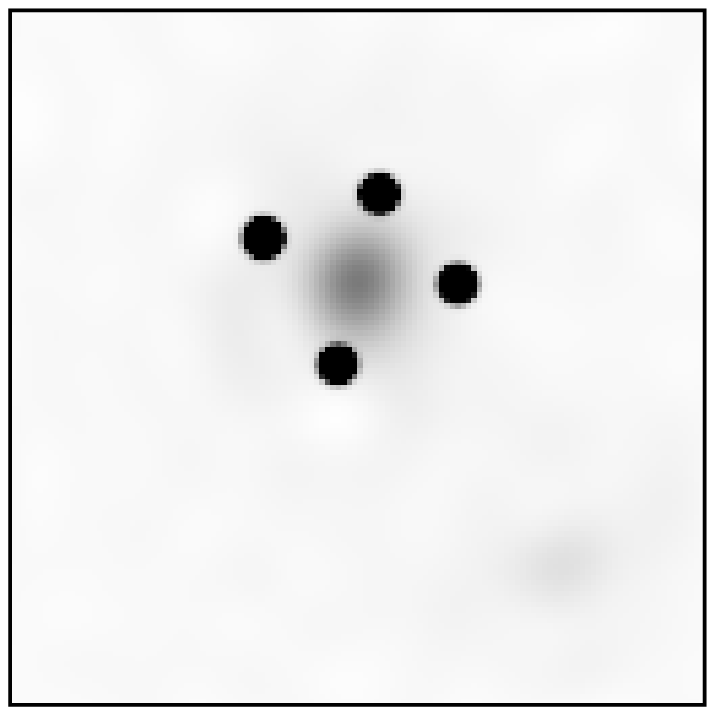}}
\put(123,42){\includegraphics[width=4cm,bb=98 98 302 302]{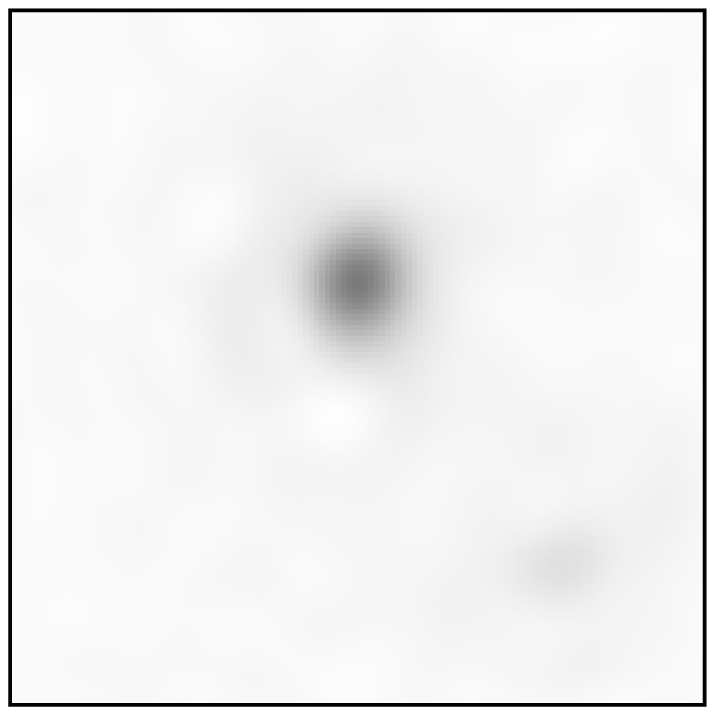}}
\put(5,48){\LARGE$r$}
\put(6,68){\large A}
\put(22,76){\large B}
\put(31,65){\large C}
\put(15,54){\large D}
\put(0,0){\includegraphics[width=4cm,bb=98 98 302 302]{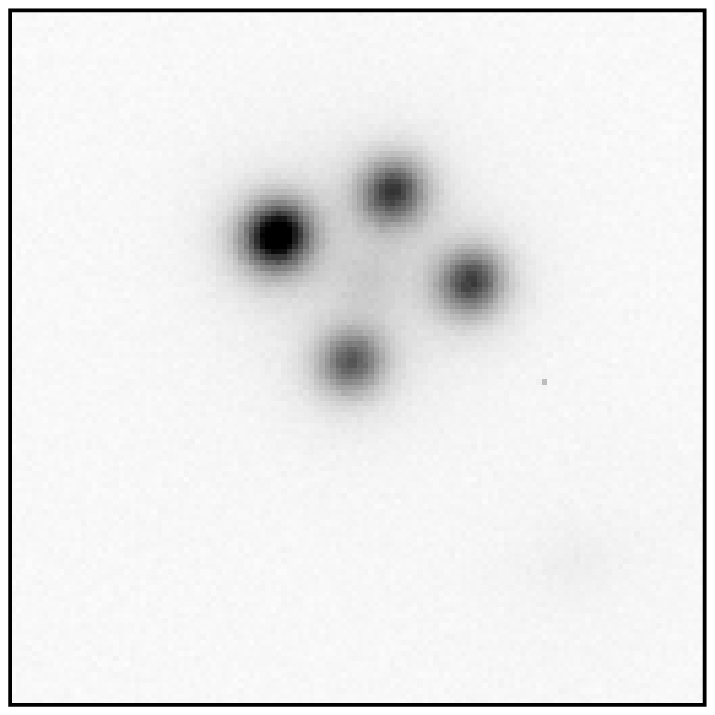}}
\put(41,0){\includegraphics[width=4cm,bb=98 98 302 302]{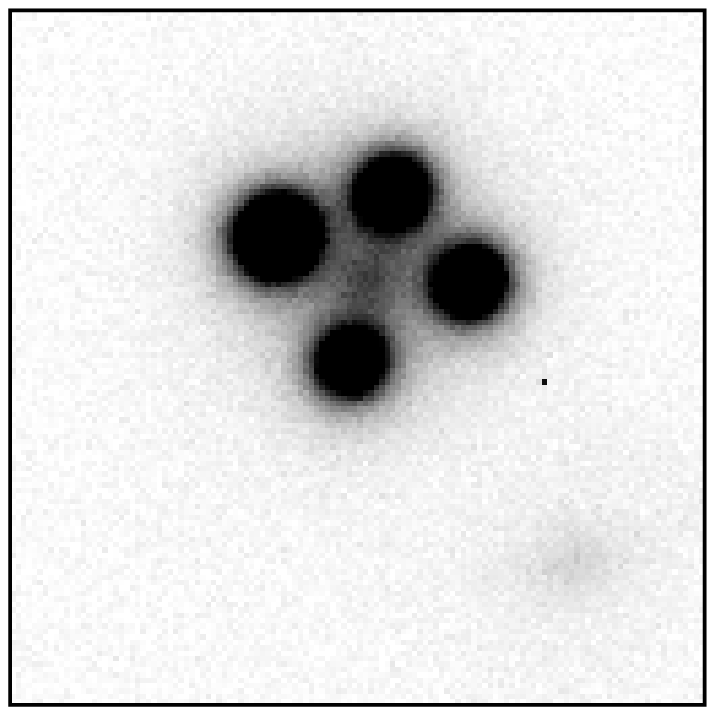}}
\put(82,0){\includegraphics[width=4cm,bb=98 98 302 302]{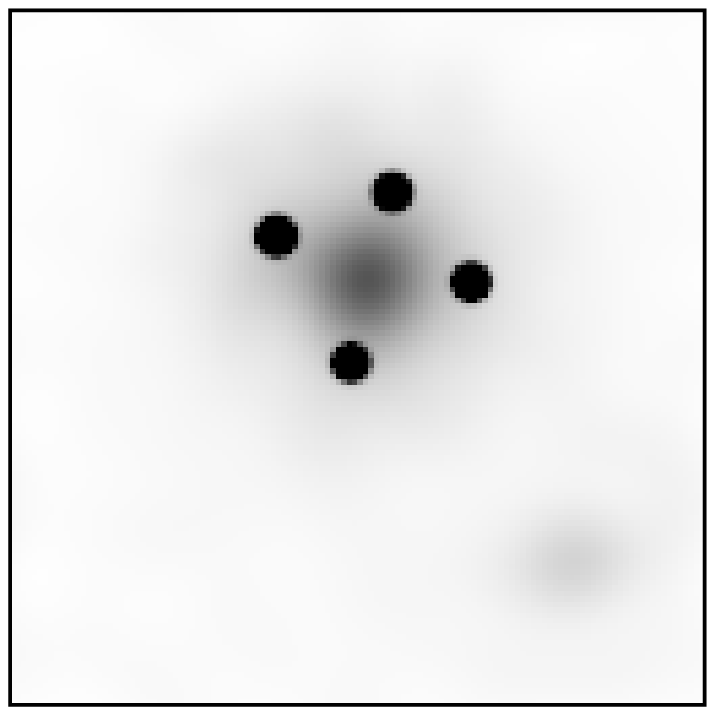}}
\put(123,0){\includegraphics[width=4cm,bb=98 98 302 302]{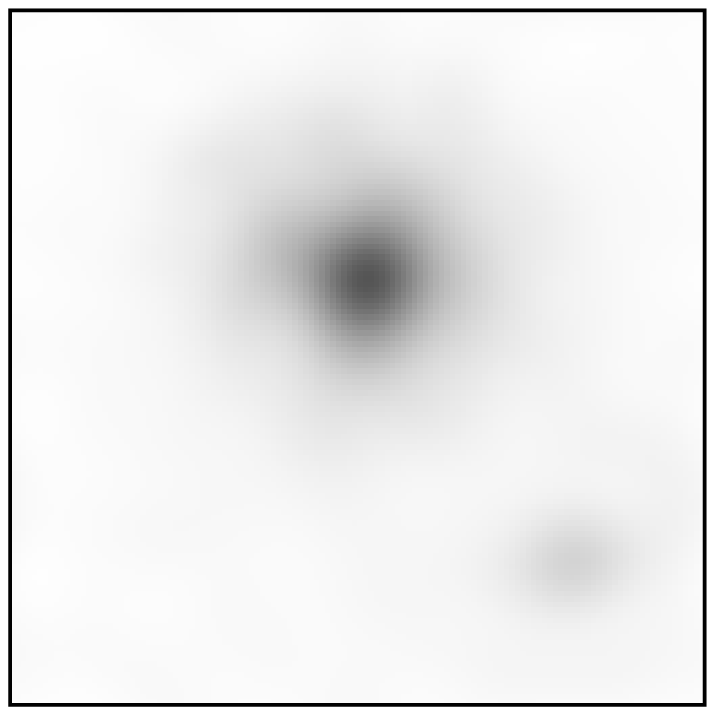}}
\put(5,6){\LARGE$i$}
\put(35.5,97.5){\large N}
\put(37,87){\vector(0,1){9}}
\put(24.5,85.9){\large E}
\put(37,87){\vector(-1,0){9}}
\end{picture}
\caption[]{Gallery of images of the quadruple QSO, with
$g$, $r$ and $i$ bands, from top to bottom. The left two panels 
show the unprocessed image with different linear cut levels.
The two right-hand panel show the deconvolved images, with and
without the quasar components. Each image measures 
$8\farcs 8\times 8\farcs 8$ in the sky; the arrows in the top left
panel are $2\farcs 0$ long. Notice the possible 
companion to the lensing galaxy to the SW.
}
\label{fig:ima}
\end{figure*}

\subsection{Imaging}

High resolution optical images were obtained at the 6.5~m Baade telescope
on 14 Dec 2001, equipped with the \emph{Magellan Instant Camera} (MagIC) 
and a 2k$\times$2k CCD. 
The image scale at the f/11 Nasmyth focus was approximately
$0\farcs0692$ per pixel.  A Shack-Hartmann wavefront sensor (Schechter
et al.\ 2002, in preparation) was used to update the focus and translation of the
secondary and the twelve most flexible elastic modes of the primary
mirror at half minute intervals.

\begin{figure}[tb]
\includegraphics[width=8cm]{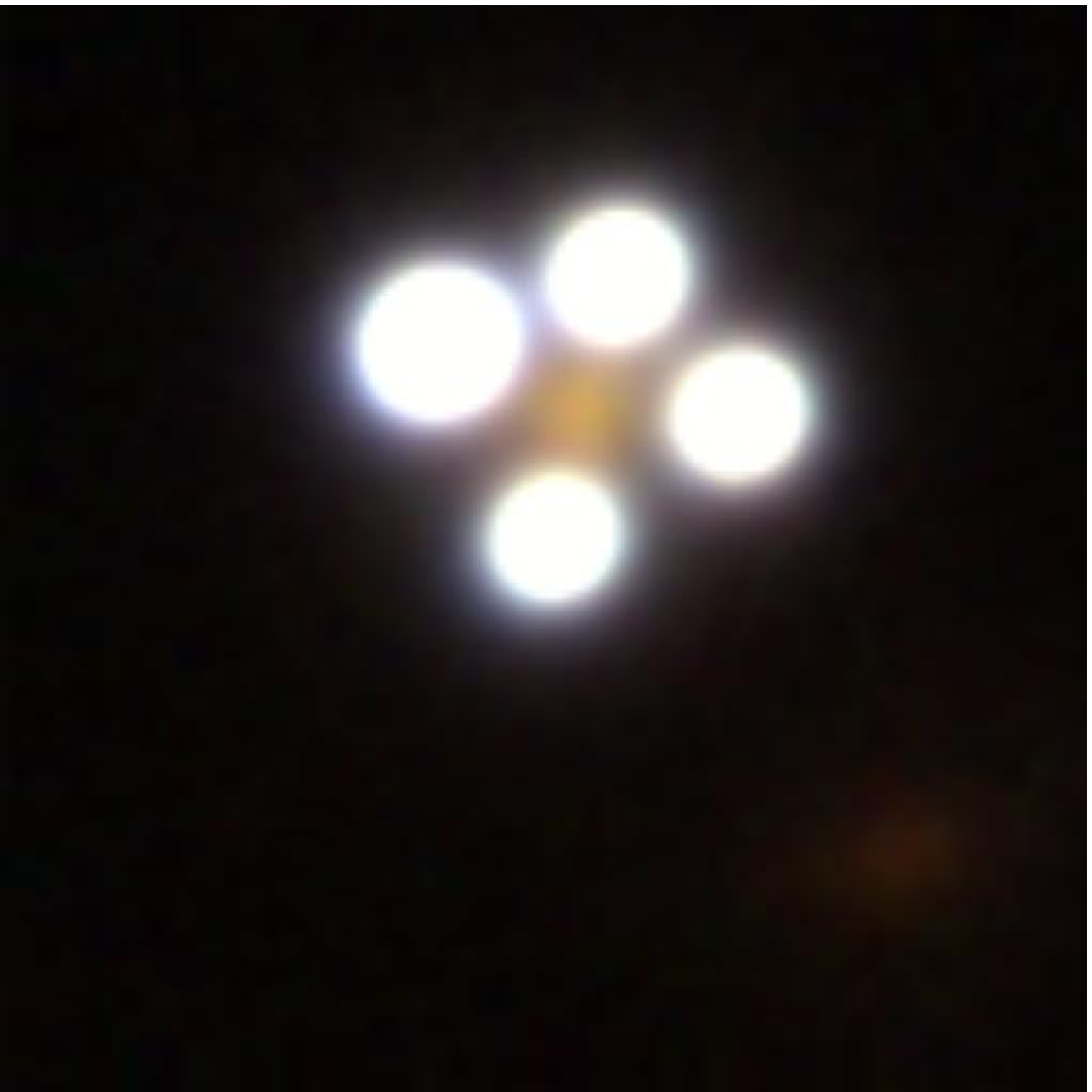} \\[0.5ex]
\includegraphics[width=8cm]{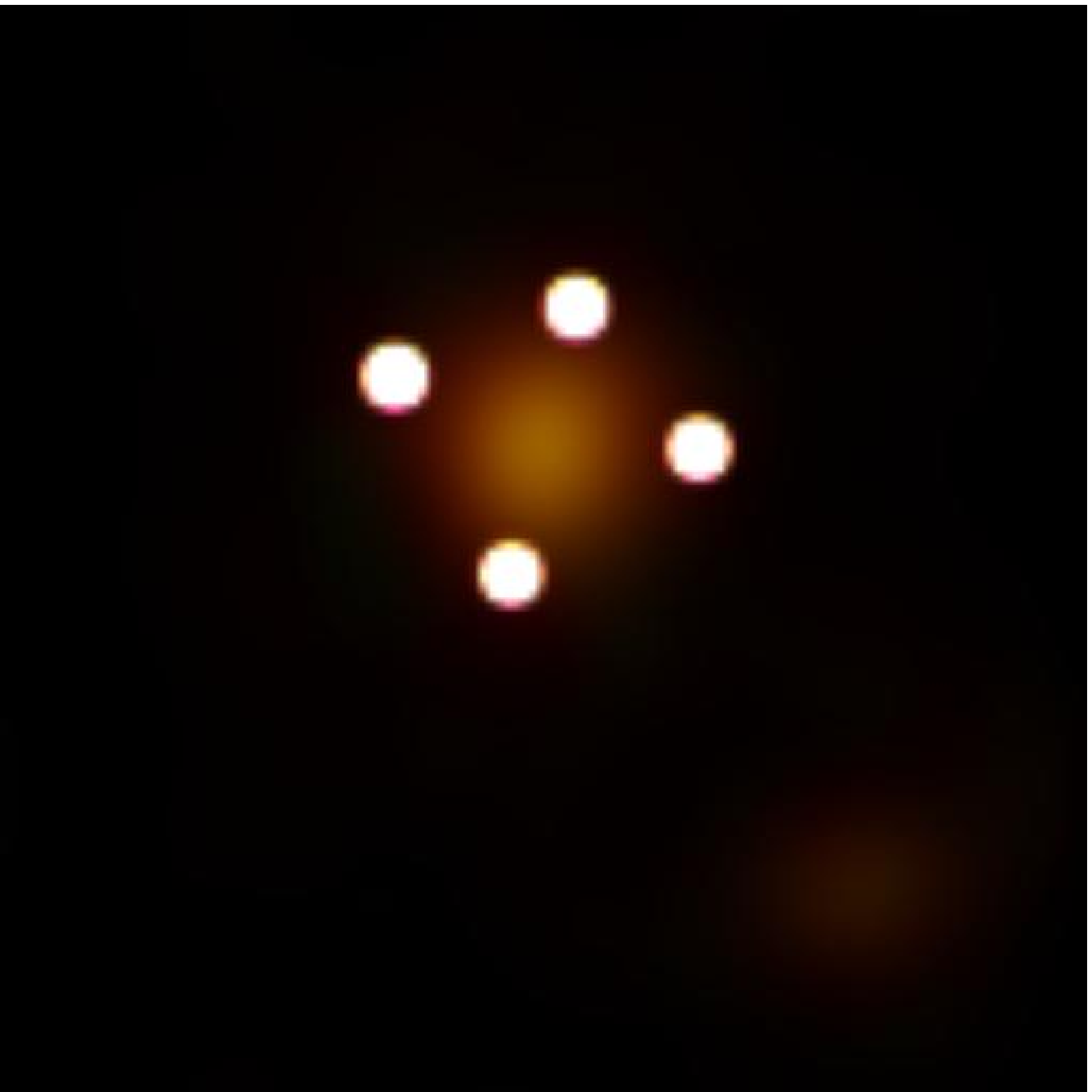}
\caption[]{Colour composite images of the 
unprocessed (left) and MCS-deconvolved (right) 
Magellan data. The RGB colour channels are represented 
by the $i$, $r$, and $g$ band data, respectively. 
Orientation as in Fig.\ \ref{fig:ima}.
}
\label{fig:rgb}
\end{figure}

After a first $r'$ band `snapshot' of 30~sec which revealed the multiple nature
of the QSO, five further images were obtained: One each in the SDSS $g'$ 
and $i'$ bands, and three in $r'$, of 300~sec exposure each
(for simplicity, we shall refer to these bands as \emph{gri} in the following).
The effective seeing was between $0\farcs 6$ and $0\farcs 8$ 
(for the $g$ image). At the scale of $0\farcs 07$/pixel,
the images were highly oversampled. In Fig.\ \ref{fig:ima} we show 
postage stamps of the QSO in the three bands. The symmetric
configuration of four bright point sources plus a fuzzy object
in the centre immediately recalls the famous `Einstein Cross'
Q~2237$+$0305 (Huchra et al.\ \cite{huchra*:85:2237}), and 
the image alone leaves little doubt that \he{} is another
case of a quadruple gravitational mirage. Given the nearly textbook
arrangement of the components in \he{}, we shall assume henceforth
that there is no reasonable alternative to the lensing hypothesis 
in this object.
A colour composite image is shown in Fig.\ \ref{fig:rgb}.

Second epoch data was secured on 15 Feb 2002, using the same 
telescope and instrument configuration as above. Three 120~sec images
in the $g$ band were taken at $1''$ seeing under photometric conditions.
The standard star sequence PG~1047+003 was observed to provide a
zeropoint for the $g$ band.

\begin{table}
\caption[]{Differential astrometry for \he{}.
Components A--D are labelled clockwise, 
starting with the brightest component A.}
\label{tab:pos}
\begin{tabular}{l.@{\mbox{\hspace{0.5em}$\pm$}\hspace{-0.75em}}.@{\hspace{2em}}.@{\mbox{\hspace{0.5em}$\pm$}\hspace{-0.75em}}.}
\hline\noalign{\smallskip}
Component & \multicolumn{2}{c}{$\Delta\alpha$} & \multicolumn{2}{c}{$\Delta\delta$} \\
          & \multicolumn{2}{c}{[arcsec]} & \multicolumn{2}{c}{[arcsec]} \\
\noalign{\smallskip}\hline\noalign{\smallskip}
A         &  \multicolumn{2}{.}{0.000} & \multicolumn{2}{.}{0.000} \\
B         &  -1.483   & 0.002 &    0.567 & 0.002\\  
C         &  -2.488   & 0.003 &   -0.589 & 0.002\\  
D         &  -0.951   & 0.001 &   -1.620 & 0.001\\  
G         &  -1.15    & 0.05  &   -0.51  & 0.05 \\
      \noalign{\smallskip}\hline
  \end{tabular}
\end{table}

\section{Analysis}

\subsection{PSF fitting and deconvolution}

In order to decompose the observed images into four point sources
plus an extended contribution from the lensing galaxy, we followed
two independent approaches. First, we performed a straightforward
PSF fitting analysis using DAOPHOT~II (Stetson et al.\ \cite{stetson:87}).
This confirmed the visual impression that components A--D (numbered 
clockwise starting from the brightest component) are indeed point-like,
but the PSF-subtracted image looked less than satisfactory especially
in the $i'$ and $r'$ bands.
The point sources were oversubtracted, and the source positions were
systematically shifted inwards, both clearly caused by the fact that 
the contribution  of the lensing galaxy has not been
taken into account in the fitting.

We have therefore chosen to apply a more sophisticated method.
The algorithm devised by Magain, Courbin \& Sohy (\cite{magain*:98:MCS};
hereafter MCS) combines multiple point source fitting with a locally
regularised deconvolution of the smooth image component. The
algorithm has been demonstrated to yield astrometrically and 
photometrically unbiased results (MCS; see also Burud et al.\ 
\cite{burud*:98:Q2237}) and is thus well-suited to our case.
We have used an `alpha release public version' of the code
to decompose each image into four point sources plus a smooth
`background' channel. The reconstructed images are shown in
the right-hand panels of Fig.\ \ref{fig:ima}, adopting a
final resolution of $0\farcs 21$ FWHM in $r$ and $i$,
and of $0\farcs 28$ in $g'$. The lensing galaxy is very
prominent in the reconstructed $i$ band image,  
fainter but clearly detectable in $r$, 
but swamped by PSF residuals in $g$. 
The non-detection in $g$ is confirmed by the DAOPHOT
analysis which also shows no significant galaxy residual 
at the expected location.

While the MCS method is probably close to optimal in its treatment 
of the point sources, we found that the exact shape of the 
reconstructed lensing galaxy depended critically on the local 
smoothing term $\lambda$ which basically is a free parameter 
in the algorithm;  slight variations of $\lambda$ (by a factor of 
$\sim 2$ or so) led to significantly different scale lengths.
We therefore reverted to a hybrid approach: 
Using the results of the MCS routine for
the point source positions and fluxes, we went back to DAOPHOT and 
performed another PSF subtraction cycle. The resulting image
of the lensing galaxy looks almost free of PSF residuals and
certainly good enough to perform a first 
reliable morphological analysis of the lensing galaxy.

\subsection{Astrometry}

Rectangular positions of the four point source components
were provided by the MCS PSF fitting/deconvolution code.
Arbitrarily adopting the brightest component A as reference
point, we list the relative positions of \he{} B--D in
Table~\ref{tab:pos}. Errors were estimated from the rms 
deviations between individual frames as well as between
individual deconvolution runs, which differ by adopting 
slightly different parameters for the smoothing scale. 
All measurements agree with each other extraordinarily well, 
with deviations from the mean of never more than a few 
milli-arcseconds, with no systematic effects visible
between the different photometric bands.

The lensing galaxy had to be treated separately, as it is
not modelled as an individual `object' by the MCS code. 
However, in the $r'$ and $i'$ bands it has a well-defined
centroid which we measured by fitting a Gaussian to the
inner region. When expressed relative to image A, the 
galaxy centroid is shifted by $\sim 0\farcs 05$ between
$i$ and $r$. We estimate the uncertainty of the galaxy 
position to be at least of this order.

In order to relate the above relative astrometry to
a commonly used reference system, we have tied
the measurements to the \emph{Digitized Sky Survey} (DSS)
astrometric solution. The quasar itself is unresolved
in the DSS images, but four nearby stars are found in all
our CCD images as well as in the DSS. We computed simple
bilinear solutions for each CCD image and obtained the
following coordinates of Component A:
R.A. = 04$^\mathrm{h}$~38$^\mathrm{m}$~14$\fs 90$,
Dec = $-12\degr$~17$'$~$14\farcs 4$ (J2000.0), with
errors defined entirely by the unknown systematics
of the DSS solution in this area.

\begin{table}
\caption[]{Differential photometry for the four point source components,
expressed as magnitudes relative to the total aperture flux in each band.
The $g$ band values are equally valid for first and second epoch data.}
\label{tab:rphot}
\begin{tabular}{lrrr}
\hline\noalign{\smallskip}
Component & \multicolumn{1}{c}{$g$} &
            \multicolumn{1}{c}{$r$} &
            \multicolumn{1}{c}{$i$} \\
\noalign{\smallskip}\hline\noalign{\smallskip}
A               & 1.10 & 1.10 & 1.11 \\
B               & 1.61 & 1.61 & 1.59 \\
C               & 1.70 & 1.67 & 1.63 \\
D               & 1.74 & 1.78 & 1.82 \\
\noalign{\smallskip}\hline
\end{tabular}
\end{table}

\begin{table}
\caption[]{Photometry for \he{} (first epoch). 
Calibrated magnitudes of 
the QSO (including galaxy) and its components are relative to 
the reference star as described in the text.}
\label{tab:phot}
\begin{tabular}{lrrrrr}
\hline\noalign{\smallskip}
Component & \multicolumn{1}{c}{$g$} &
            \multicolumn{1}{c}{$r$} &
            \multicolumn{1}{c}{$i$} &
            \multicolumn{1}{c}{$g-r$} &
            \multicolumn{1}{c}{$r-i$} \\
\noalign{\smallskip}\hline\noalign{\smallskip}
QSO total       &    17.89 & 17.15 & 16.53 &    0.74 & 0.62 \\[0.5ex]
A               &    19.00 & 18.44 & 17.95 &    0.56 & 0.49 \\
B               &    19.51 & 18.95 & 18.43 &    0.56 & 0.52 \\
C               &    19.60 & 19.01 & 18.47 &    0.59 & 0.53 \\
D               &    19.64 & 19.10 & 18.66 &    0.54 & 0.44 \\[0.5ex]
G ($\infty$)    &          & 19.16 & 18.05 &         & 1.11 \\
G ($0\farcs 7$) & $>$22.70 & 20.91 & 19.82 & $>$1.80 & 1.08 \\[0.5ex]
Ref.\ star      &    17.80 & 16.56 & 15.98 &    1.24 & 0.58 \\
\noalign{\smallskip}\hline
\end{tabular}
\end{table}

\subsection{Photometry} \label{sec:phot}

Relative photometry of components A--D is directly provided 
on output by the MCS code. Even without any calibration it is 
clear that all components have very similar colours. In 
Table~\ref{tab:rphot} we present the fractional contributions
of each component to the total QSO flux (without lensing galaxy),
expressed in magnitudes. Between the photometric bands, 
the contributions of components A and B show deviations of
no more than $\pm 0.01$~mag, while components C and
D show marginally significant deviations but are still very 
nearly independent of wavelength, within $\pm 0.05$~mag.
Note that the numbers for the $g$ band are identical 
between first and second epoch to within $\pm 0.01$~mag.

For measuring the brightness of the lensing galaxy we 
constructed PSF-subtracted images with DAOPHOT, but using
the relative astrometry of the QSO components from the MCS
deconvolution. The resulting galaxy-only images look quite
clean and devoid of strong PSF subtraction residuals, except
for the $g$ band where we regard the lensing galaxy to be undetected, 
because of the significant PSF residuals in that image. 
We quote two values for the galaxy flux: 
one obtained by integrating to infinity and 
a second through a small aperture of 10 pixels radius (0\farcs 7).
The small aperture was included because within this radius,
residual PSF contamination should be close to negligible.
While a significant fraction of the total galaxy \emph{flux} will 
be missed this way, the galaxy \emph{colours} should be rather reliable. 
For the $g-r$ colour, the formally measured $g$ band aperture flux
of the galaxy is treated as an upper limit.

Since the first epoch data were obtained under nonphotometric conditions, 
we have tied the $g$ band photometry of the QSO to a nearby star,
visible within the CCD field of view and located at
R.A. = 04$^\mathrm{h}$~38$^\mathrm{m}$~12$\fs 9$,
Dec = $-12\degr$~17$'$~$52''$ (J2000.0). We found this star to
have $g = 17.80$ in the second epoch data. Assuming the star
to be non-variable, the total (large aperture) magnitude of the
QSO is $g = 17.89$ a the first epoch and $g = 17.71$ at the
second epoch, with uncertainties of at most $\pm 0.02$~mag.
There is thus evidence for significant variability of the QSO.
However, as stated above, the fractional contributions
of the individual QSO components to the total $g$ band flux
are identical in first and second epoch data. 
We return to this issue below.

As the second epoch provided only calibrated $g$, 
but no $r$ and $i$ data,
we have worked out at least approximate zeropoints for
these latter bands as follows.
We found that the above reference star has an entry 
in the database of digital objective prism spectra of the
Hamburg/ESO Survey (Wisotzki et al.\ \cite{wisotzki*:00:HES3}).
Christlieb et al.\ (\cite{christlieb*:01:DA}) showed that these 
spectra can be used to obtain ($B-V$) colours of normal stars 
with an accuracy of better than 0.1~mag. Following
their recipe, we estimate $B-V \simeq 1.1$ for our star
and then used a matching model atmosphere spectrum%
  \footnote{obtained from R.~Kurucz' web site at \\ 
  \texttt{http://cfaku5.harvard.edu/}}
to estimate its colours in the SDSS system to
$g-r \simeq 1.24$, and $g-i \simeq 1.82$. 
Combining the second epoch $g$ band calibration with 
the estimated intrinsic colours of the reference star,
we could thus tie the first epoch $r$ and $i$ band data to 
an external photometric system.
The uncertainty of this `calibration' procedure 
is considerable, leading to zeropoint errors of 
$\sim 0.15$~mag for $r$ and $\sim 0.2$~mag for $i$ 
(assuming a standard error of $\pm 0.1$~mag in $B-V$ 
of the reference star). 
However, the errors in $r$ and $i$ are strongly correlated, 
and the zeropoint uncertainty in $r-i$ colour
is only $\sim 0.06$~mag.
Results for the final calibrated photometric results 
are  collated in Table~\ref{tab:phot}.

\begin{figure}[tb]
\includegraphics[width=8.8cm,bb=72 84 341 344,clip]{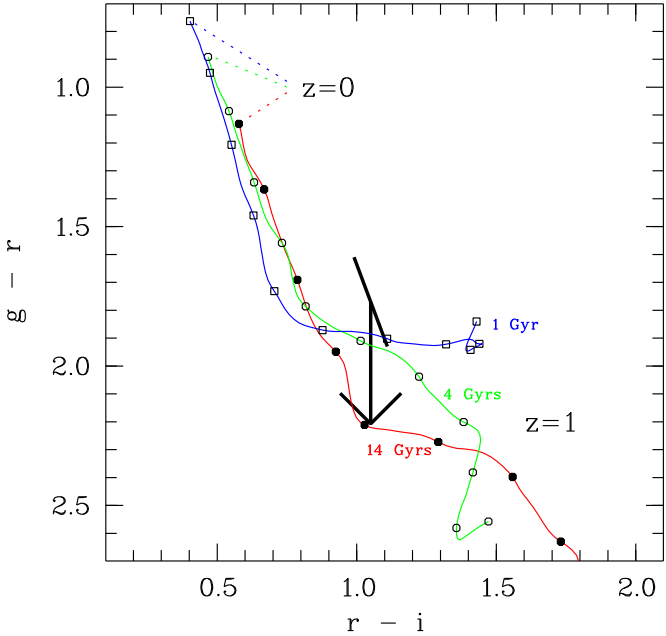}
\caption[]{$r-i$ vs.\ $g-r$ colour variations of single stellar 
population models, as a function of redshift.
Steps of $\Delta z = 0.1$ are denoted by the small tickmarks
(filled symbols: population age 14~Gyrs; open circles: 4~Gyrs;
open squares: 1~Gyr).
The arrow+bar symbol represents the measured $r-i$ and 
upper limit on $g-r$ of the lensing galaxy. The inclination
of the error bar indicates the correlated errors in the two 
colours.}
\label{fig:photoz}
\end{figure}

\begin{figure}[tb]
\includegraphics[width=8.8cm,bb=77 87 342 286,clip]{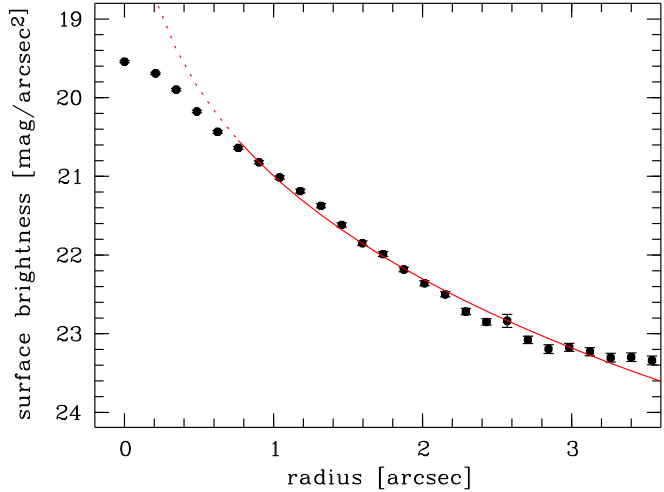}
\caption[]{Radial surface brightness profile
of the lensing galaxy, measured in the PSF-subtracted $i$ band image. 
The solid line shows the best-fit de Vaucouleurs profile (inner regions
extrapolated as shown by the dotted line). 
}
\label{fig:galprof}
\end{figure}

\section{Discussion}

\subsection{Properties of the lensing galaxy}

Lacking direct spectroscopic evidence, we have estimated the
redshift of the lensing galaxy using its measured colours.
Figure~\ref{fig:photoz} shows
the redshift-colour dependence of three simple galaxy models, 
represented as single stellar populations (SSPs)
with ages of 1, 4, and 14 Gyrs, respectively. 
(For the limited redshift baseline covered, we ignored
passive evolution.) We used the recent 
solar metallicity models from 
Jimenez et al.\ (\cite{jimenez*:00:PS}),
kindly provided by R. Jimenez.
The bar+arrow symbol denotes the measurement 
of the galaxy inside the small ($0\farcs 7$) aperture,
with the width and inclination of the bar reflecting 
chiefly the uncertain calibration (recall that the errors
in $r-i$ and $g-r$ are correlated).

This plot suggests that the redshift is probably around
$z=0.4$--0.5, the best estimate depending on the
unknown intrinsic spectrum of the galaxy. The youngest
(1~Gyr) SSP model yields the highest redshift ($z_l=0.55$)
and is only marginally consistent with our upper limit for
the $g$-band flux. With the more reasonable assumption of 
an old stellar population dominating the SED of the lensing galaxy,
the best redshift estimate drops to $z=0.41$ for a 14~Gyrs
old SSP. Because the 14~Gyrs track bends sharply in exactly 
this part of the colour-colour diagram, the corresponding
formal error is highly asymmetric. Adopting a $2\sigma$
confidence limit, we predict a redshift in the interval 
$0.3 < z_l < 0.5$; other arguments detailed below will show 
that it is more likely to be near the lower end of that range.
Of course, two colours are insufficient to distinguish an evolved
population from a younger one that involves dust extinction, 
but a very dusty lensing galaxy can probably be ruled out from
the fact that the four QSO images have such similar colours.

For the formally best redshift estimate of $z_l=0.41$, the
absolute magnitude $M_V$ of the lensing galaxy can be converted
directly from the $i$ band measurement without any $K$ correction.
We obtain $M_V = -23.8$ for $H_0 = 65$~km s$^{-1}$ Mpc$^{-1}$,
$\Omega_m = 0.3$, $\Omega_\Lambda = 0.7$, about a factor of 5
brighter than an $L^\star$ field galaxy. This would drop
to $M_R = -23.0$ or $\sim 2L^\star$ for $z_l=0.3$.

The reconstructed $i$ band image, without the point source components,
allows us to investigate the gross morphological properties
of the lens. Figure \ref{fig:galprof} shows the radial surface
brightness profile, measured in concentric annulli of 2 pixels width
($0\farcs 138$).
Inside of $\sim 0\farcs 6$, the profile is blurred by the seeing,
but outside this radius it is clear that the galaxy 
follows approximately a de Vaucouleurs $r^{1/4}$ law. The best-fit 
scale length is $r_e = 2\farcs 0$ corresponding to $(12\pm 3)$~kpc 
at $z_l = 0.41$ (redshift uncertainty not included in error estimate).
Notice that the galaxy is nearly perfectly round,
with isophotes having $1 - b/a < 0.1$ everywhere.

A possible companion is found $5''$ to the SW of the 
lensing galaxy, at a position angle of $217^\circ$. It
is visible in all bands, although the detection in $g$ is only marginal.
Its colours, measured over an aperture of radius 10~pixels or $0\farcs 7$,
are $r - i \simeq 1.4\pm 0.1$ and $g-r \simeq 1.6\pm 0.2$ (again with
strongly correlated errors). The colours are thus consistent with the
`companion' being at the same redshift as the galaxy, but without 
spectroscopic confirmation this conclusion must remain speculative.
The total magnitude of the companion is $i = 19.6$,
corresponding to $M_V = -22.3$ or $\sim L^\star$ for $z=0.41$.


\subsection{The lensing potential}

\subsubsection{Qualitative features}

The image configuration is roughly diamond shaped (indicating perhaps
a diamond in the rough), and nearly bilaterally symmetric with respect
to the AC diameter.  The AC diameter is a factor of 1.14 times larger
than the BD diameter.  Taking the potential of the lensing galaxy to
be roughly isothermal, this implies that the source of the quadrupole
moment, whether due to a bar or a tidal perturber, lies along the BD diameter
(e.g.\ Witt, Mao and Schechter \cite{witt*:95:UML}).  
The strength of the shear should
be roughly half (AC -- BD)/(AC + BD), or $\sim 0.07$.  The strength of
the isothermal sphere model ought to be half the average of the two
diameters, or 1.203 arcseconds.

\subsubsection{Quantitative aspects}

A model which works quite well for many lenses is a singular
isothermal sphere with an external shear, as might be generated by the
tide from a neighboring galaxy or cluster.  The two dimensional
projected potential (e.g. Kochanek \cite{kochanek:91:ILGS}) is given by
\begin{equation}
\psi_{2D}(\vec \theta) = br + {\gamma \over 2} r^2 \cos(\phi -
\phi_\gamma)
\end{equation}
where $b$ is the diameter of the isothermal sphere in arcseconds, $r$
and $\phi$ are the radial and angular parts, respectively, of the
angular position $\vec \theta$ on the sky with respect to the galaxy,
and $\phi_\gamma$ is the orientation of the shear on the sky,
measured E of N.  Given our sign convention, a bar or tidal perturber
would be at right angles to $\phi_\gamma$.  In addition to the three
explicit parameters, the position of the source adds two more
parameters.

Fitting this 5 parameter model to the positions (but not the
fluxes) of the four images we find that the model gives rms errors of
$\sim 0\farcs05$.  The rms residuals improve considerably if we take
the position of the lensing galaxy to be free, adding two additional
parameters.  The predicted galaxy position is then some 
$\sim 0\farcs06$ from the observed position, consistent with our measurement
errors.  We find $b=1\farcs209$, $\gamma = 0.074$ and $\phi_\gamma =
76.5^\circ$, with a source position ($\Delta \alpha, \Delta\delta$) =
($-0\farcs0423$, $-0\farcs0076$). The source of the tide is predicted to
lie at P.A. $-13.5^\circ$ or $166.5^\circ$.

\begin{table}
\caption[]{Observed and predicted quantities for the images and
the lens.  Positions are in arcseconds, relative to the observed lens centre.
$\mu$ is the magnification factor, $t_i$ is a quantity related to
the time delay (see text, Sect.~\ref{sec:td}).}
\label{tab:modpos}
\begin{tabular}{lrrrrrr}
\hline\noalign{\smallskip}
          & \multicolumn{2}{c}{observed}  
          &  \multicolumn{2}{c}{predicted}
          &
          &                               \\
          & \multicolumn{1}{c}{$\Delta\alpha$} 
          & \multicolumn{1}{c}{$\Delta\delta$}
          & \multicolumn{1}{c}{$\Delta\alpha$} 
          & \multicolumn{1}{c}{$\Delta\delta$} 
          & $\mu$                          
          & $t_i$ \\
\noalign{\smallskip}\hline\noalign{\smallskip}
A              &    1.150 &    0.510 &    1.150 &    0.509 &    7.4 & $-$0.798 \\
B              & $-$0.333 &    1.077 & $-$0.334 &    1.077 & $-$8.2 & $-$0.725 \\
C              & $-$1.338 & $-$0.079 & $-$1.339 & $-$0.078 &    7.7 & $-$0.790 \\
D              &    0.199 & $-$1.110 &    0.201 & $-$1.110 & $-$5.0 & $-$0.636 \\
G              &    0.000 &    0.000 & $-$0.036 & $-$0.047 &        &        \\
\noalign{\smallskip}\hline
\end{tabular}
\end{table}

In Table~\ref{tab:modpos} we give the observed and predicted positions
for the four images, as well as the predicted magnifications, $\mu$.
Negative magnifications indicate the parity flip associated with
saddlepoints of the time delay function.  Components A and D are both
observed to be brighter than predicted by $\sim 0.5$ mag relative to
components B and C.

For an isothermal sphere, the lens strength (measured in radians) is
given by
\begin{equation}
b = {D_{ls} \over D_{os}} {4 \pi \sigma^2 \over c^2}
\end{equation}
(Narayan and Bartelmann \cite{nara+bart:99:GL}) where $\sigma$ is the line of sight
velocity dispersion of the isothermal sphere and where $D_{os}$ and
$D_{ls}$ are angular diameter distances, from observer to source and
lens to source, respectively.  Taking $z_l = 0.41$ we get $\sigma =
252$ km/s.  The Faber-Jackson relation would predict a substantially
larger dispersion for a $5L^\star$ galaxy.  At $z_l = 0.30$, we get $\sigma
= 237$ km/s, more nearly in line with the expectation for a $2L^\star$
galaxy.

The shear, $\gamma = 0.074$, while not especially large for a
quadruple lens, would call for an E3--E4 mass distribution if the flattening 
of the galaxy were the source of the
quadrupole moment (Keeton et al.\ \cite{keeton*:98:OPGL}).
As the flattening of the
light is unmeasurably small, this suggests that the quadrupole moment
is indeed due to an external tide.  In several quadruple systems the
source of the tide, a cluster or group of galaxies, is clearly visible
at right angles to $\phi_\gamma$ (Kundic et al.\ \cite{kundic*:97:TD};
Schechter et al.\ \cite{schechter*:97:PG1115}; 
Kneib et al.\ \cite{kneib*:98:MCL}).  In the present case the nearest
galaxy, at P.A. $217^\circ$, is not at the expected position angle.
There are several more galaxies in the neighborhood, again with no
obvious concentrations at the expected position angle.

\subsubsection{Time delay} \label{sec:td}

The gravitational time delay function, which gives the additional
travel time for an image at position $\vec \theta_i$, is given by
\begin{equation}
\tau_i = {1+z_l \over c} {D_{ol} D_{os} \over D_{ls}}
{\left[{1 \over 2}(\vec\theta - \vec \beta)^2 - \psi_{2D}(\vec \theta)\right]
\over (206265)^2}
\end{equation}
where $\vec \beta$ is the position of the source, and where we have
explicitly included the transformation from arcseconds to radians.  
As the redshift of the lens is poorly known, in Table~\ref{tab:modpos} we
give only the square bracketted quantity, labelled $t_i$, which has
units of (arcsecond)$^2$.  We see that components A and C lead,
followed by saddlepoints B and D.  The longest delay, $\tau_D -
\tau_A$ is $11\fd8$ for $z_l = 0.41$  or $7\fd9$ for $z_l = 0.30$.  
Predicted time delays are quite sensitive to the
position of the lensing galaxy for symmetric lenses.  The time delays
predicted using the observed position for the lensing galaxy are 15\,\%
smaller.


\subsection{Variability}

In Sect.\ \ref{sec:phot} we demonstrated that the
total QSO flux in the $g$ band varied by almost 20\,\%
within a timespan of 2 months. A comparison between
our 2001/2002 CCD photometry and the original photographic
discovery data reveals additional, albeit circumstantial, 
evidence for variability: On the UKST image of the Digitized 
Sky Survey, calibrated by photometric sequences taken
in the course of the Hamburg/ESO Survey, we measure the
QSO to have $B_J = 16.2\approx g$ at epoch 1984.9.
On the spectral discovery plate obtained with the ESO
Schmidt telescope (epoch 1996.9), the object has only
$B_J = 17.2$, calibrated against the same photometric 
sequence. While the HES photographic photometry is well-defined
only for point sources (with a global rms accuracy
of $\sim 0.15$~mag), which might lead to a photometric
bias for multiply imaged QSOs, inspection of the scan data 
revealed the quasar to appear perfectly unresolved.
We conclude that the differences between photographic and 
the more recent CCD measurements are most probably significant.

For the two Magellan epochs Dec~2001 and Feb~2002, we can
state that the variability is almost certainly intrinsic 
to the QSO and not caused by microlensing. This is given
by the fact that the flux ratios of the four components
have remained unchanged to within 1\,\%, despite an almost
20\,\% variation of total flux. Microlensing-induced flux
amplification would always affect each image differently,
which is in contradiction to the observations.
On the other hand, the short expected time delay means
that intrinsic variations (assumed to be much slower) 
will show up quasi-simultaneously in all four images. 

While we can exclude microlensing as the cause of variability
between Dec~2001 and Feb~2002, this cannot be said about the 
origin of the much larger variations between 1986 and today.
In fact, the amplitude of more than 1~mag might even favour
microlensing, as such amplitudes are rare for intrinsic
variations of normal, non-OVV QSOs 
(Hook et al.\ \cite{hook*:94:VOSQ}; 
Cristiani et al.\ \cite{cristiani*:96:OVQ}).
From the extensive monitoring campaigns of
the Einstein Cross (Irwin et al.\ \cite{irwin*:89:Q2237};
Wozniak et al.\ \cite{wozniak*:00:Q2237}; OGLE web site), 
on the other hand, it is known that microlensing amplifications of 
more than one magnitude are possible and in fact happening.

\section{Conclusion}

The new quadruple QSO \he{} is an almost textbook example for
gravitational lensing, with its four nearly identical components
arranged symmetrically around a luminous early-type galaxy.
Unlike most other known quadruple systems, photometric monitoring
of this object should be relatively easy even in mediocre seeing
conditions, because of its wide image separations. Furthermore,
its location in the sky makes it accessible to both Northern
and Southern observatories.

Owing to its symmetry, the time delay is expected to be short,
and accurate measurement of differential time delays might 
therefore be difficult unless the QSO should prove to be variable
on very short timescales. This could limit the usefulness of
the object for cosmological purposes, but at the same time
it makes it an attractive target for microlensing studies,
because of the relative ease to separate intrinsic and 
microlensing-induced variations. Notice that compared to
the Einstein Cross Q~2237$+$0305, the higher lens redshift 
in \he{} implies a $\sim 10\times$ lower projected transverse 
velocity and hence a much longer characteristic timescale for 
high-amplification event from microlensing 
It will therefore be easier to obtain a well-sampled lightcurve,
but unfortunately events will be rarer and
take much longer to get covered.

We have presented evidence that the QSO experiences
substantial flux variations on time scales of months and years.
Whether microlensing could have a contribution in these
variations is not yet clear, but it can already be said with
certainty that monitoring of \he{} will be a promising task.

\begin{acknowledgements}
The Hamburg/ESO Survey was supported as ESO key programme 
02-009-45K (145.B-0009). 
MagIC was built with help from a gift by 
Raymond and Beverly Sackler to Harvard University 
and a US NSF grant, AST99-77535, to MIT.
PLS gratefully acknowledges the support of a fellowship from the
John Simon Guggenheim Foundation.
\end{acknowledgements}


\end{document}